\documentclass[11pt]{article}
\usepackage[T1]{fontenc}
\usepackage{lmodern}
\usepackage[letterpaper,margin=0.92in]{geometry}
\usepackage{amsmath,amssymb}
\usepackage{booktabs,array}
\usepackage{graphicx}
\usepackage{microtype}
\usepackage{xcolor}
\definecolor{gcblue}{HTML}{214D82}
\definecolor{gcteal}{HTML}{076761}
\definecolor{gcpurple}{HTML}{633E87}
\usepackage[colorlinks,linkcolor=gcblue,citecolor=gcteal,urlcolor=gcblue]{hyperref}
\hypersetup{
  pdftitle={GC-MoE K3: A Green-Context Role Graph for Latent Mixture-of-Experts Decode},
  pdfauthor={Yong Li, Haoyu Wang, Kan Liu, Tao Lan},
  pdfsubject={Expert-parallel Kimi K3 inference on datacenter Blackwell},
  pdfkeywords={mixture of experts, Kimi K3, Green Contexts, expert parallelism, Blackwell}
}
\newcommand{\Kthree}{GC-MoE K3}
\newcommand{\bfround}{\operatorname{bf16}}
\newcommand{\rms}{\operatorname{RMSNorm}}
\newcommand{\US}{\ensuremath{\mu\mathrm{s}}}
\newcolumntype{L}[1]{>{\raggedright\arraybackslash}p{#1}}

\title{\textbf{GC-MoE K3}\\
A Green-Context Role Graph for Latent Mixture-of-Experts Decode}
\author{Yong Li \quad Haoyu Wang \quad Kan Liu \quad Tao Lan\\
Alibaba Inc.\\
{\small\texttt{\{brucelee.ly, why492961, liukan.lk, tao.lant\}@alibaba-inc.com}}}
\date{}

\begin{document}
\maketitle

\begin{abstract}
Kimi K3 combines a latent routed mixture of experts with full-width shared experts. A complete feed-forward evaluation must score and select experts, project tokens into the latent space, exchange tokens across expert-parallel ranks, compute the routed and shared branches, and normalize and project the routed sum before joining the branches. We present \Kthree{}, a datacenter Blackwell implementation that schedules this entire evaluation as a CUDA Graph across two Green Contexts. Five base roles split input projection, routing, dispatch, expert computation, and finalization. Dispatch can fill the routed input ring while the compute partition evaluates shared experts. Finalize waits for the expert owners needed by each token and overlaps routed reduction and RMSNorm with loading the output projection weights. Conditional roles reuse the communication partition for routed computation and reduce split-$K$ projection partials. On GB200 at EP8, EP16, and EP32, the operator is faster at all 24 measured configurations than the paired SGLang-style reference (mean speedup $1.1324\times$). A comparison with historical vLLM-style measurements gives $1.3651\times$ across the same 24 shapes. Both latency comparisons time single-layer graph replay on synthetic K3-shaped inputs with a residual prefix. The separate paired numerical suite passes all 142 cases and 2{,}936 rank checks.
\end{abstract}

\section{Introduction}

Expert parallelism (EP) distributes the weights of routed experts across ranks. For each token, the layer must discover its expert owners, move the activation to those owners, and return their outputs for a weighted sum. Prior GC-MoE work separates token movement from expert matrix multiplication with CUDA Green Contexts, first on Hopper and datacenter Blackwell and then in a different form on SM120 \cite{gcbase,gcsm120}. That work motivates using GPU resources separately for communication and arithmetic. K3 changes where this overlap begins and ends. The router and shared experts read the full hidden state. Routed experts read a narrower projection. Each token must combine its selected expert responses before normalization and up projection.

\Kthree{} moves the timed boundary from a pre-routed activation/route tuple to the entire K3 feed-forward sublayer. The input is the BF16 output of post-attention normalization and its output is the projected routed result plus the shared result, optionally with a residual prefix. The fifth role in the graph combines expert responses and streams normalized rows into the output projection. Two conditional roles reclaim communication SMs for routed work and reduce split-$K$ projection partials. We evaluate the complete operator over 24 GB200 configurations and report the provenance of both baseline series.

\section{K3 workload and comparison paths}
\label{sec:workload}

\subsection{The complete K3 feed-forward sublayer}

Let $X\in\mathbb{R}^{T\times7168}$ contain the normalized post-attention FFN input on one rank. The router applies a $7168\to896$ linear map. It selects the top 16 experts by the sum of a sigmoid score and an expert-specific correction bias, then normalizes the selected sigmoid scores before any routed scaling. A separate $7168\to3584$ projection gives latent activations $Z$ to all selected routed experts. Each routed expert transforms a row of $Z$ through two input projections, a SiTU gate, and one output projection in a $3584\to3072\to3584$ MLP. For gate value $g$ and up-projection value $u$, the evaluated SiTU is
\begin{equation}
  \phi(g,u) = \bigl[4\tanh(g/4)\,\sigma(g)\bigr]\bigl[25\tanh(u/25)\bigr].
\end{equation}
Each of the two shared experts evaluates a $7168\to3072\to7168$ SiTU MLP on $X$. The implementation concatenates their intermediate channels into one 6144-wide shared MLP, so its output projection sums both contributions. Write $S_t$ for their sum at token $t$, $\mathcal E_t$ for that token's Top16 set, $a_{t,e}$ for its route weights, and $f_e$ for a routed expert. Let $c$ be the routed scaling factor (the benchmark default is 1). In real arithmetic the output is
\begin{equation}
  R_t=\sum_{e\in\mathcal E_t}a_{t,e} f_e(Z_t),\quad
  U_t=\rms(cR_t)W_{\rm up}^{\top},\quad Y_t=U_t+S_t+P_t .
  \label{eq:semantic}
\end{equation}
$P_t$ is an optional full-width residual prefix, omitted when absent. RMSNorm normalizes the combined row across its 3584 channels, then applies the learned per-channel weight. The harness uses $\epsilon=10^{-6}$. $W_{\rm up}$ is shared by all routed experts. With 896 routed experts, EP8, EP16, and EP32 assign 112, 56, and 28 complete routed experts per rank, respectively. The shared experts are local and replicated in our benchmark.

\subsection{SGLang and vLLM}
\label{sec:baselines}

\begin{figure}[!htbp]
  \centering
  \includegraphics[width=\linewidth]{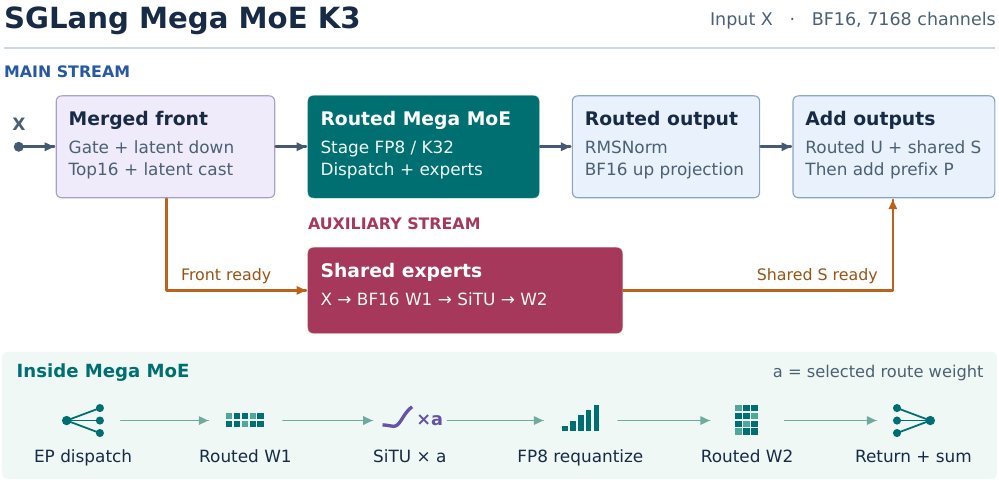}
  \caption{SGLang-style schedule for the measured decode shapes. The auxiliary stream starts the shared MLP after the merged front end, reading the original input $X$. The main stream stages the latent input, runs routed Mega MoE, then normalizes and projects the routed sum. The final addition waits for the shared result $S$. Amber arrows indicate cross-stream ordering.}
  \label{fig:sglang-flow}

  \vspace{12pt}
  \includegraphics[width=\linewidth]{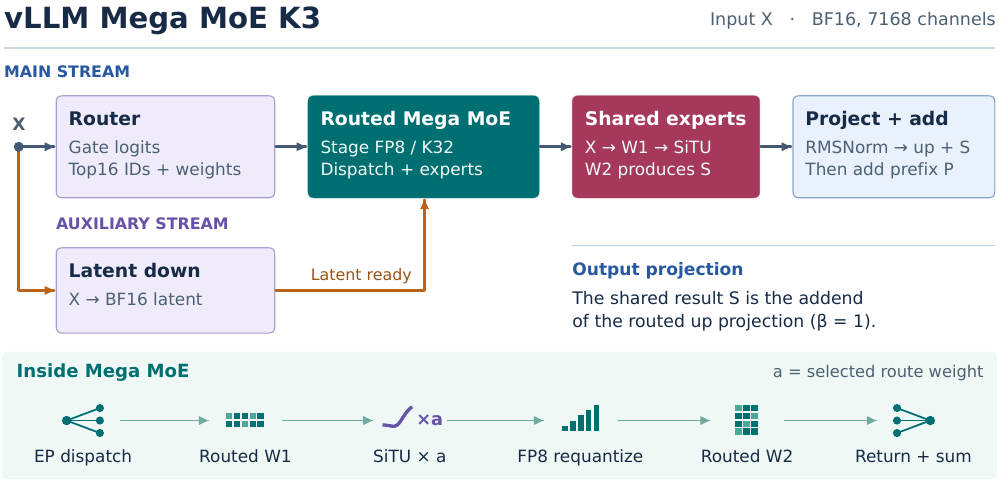}
  \caption{vLLM-style schedule for the same shapes. Gate and Top16 on the main stream can overlap the auxiliary latent down projection. The streams join before staging. Mega MoE and the shared MLP then run in sequence on the main stream. The BF16 up projection accumulates into $S$, and a separate addition applies the residual prefix $P$.}
  \label{fig:vllm-flow}
\end{figure}

Both framework paths use a DeepGEMM Mega MoE call for the latent routed computation and compute the full-width shared branch separately \cite{sglang,sgldg,vllm,vllmdg}. Figures~\ref{fig:sglang-flow} and~\ref{fig:vllm-flow} show where each benchmark wrapper forks and rejoins its auxiliary stream.

\paragraph{SGLang.}
SGLang selects among merged, two-stream, and independent front projections by device and shape. On every measured shape, the wrapper uses one BF16-weight GEMM for router logits and the latent down projection. A Triton epilogue selects Top16 and casts the latent output to BF16. A second Triton step packs FP8 activations, K32 scales, and routes for \texttt{sgl-deep-gemm} 0.2.8. The auxiliary shared MLP uses shape-dependent FlashInfer, TGV, or PyTorch projections. By default the wrapper caps Mega MoE at two fewer streaming multiprocessors (SMs) than the device total.

\paragraph{vLLM.}
The measured wrapper follows the stock BF16 schedule in Figure~\ref{fig:vllm-flow}. It uses the pinned vLLM DeepGEMM fork for routed experts and local Triton for Top16, staging, shared SiTU, and routed RMSNorm. The small-batch BF16 gate is optional. The archived measurements do not identify the gate used at each shape. An alternative K3.1 serving schedule for eligible captured TP1+EP batches packs shared W1 with the latent down projection and runs shared W2 on a side stream.

\newpage
\begin{table}[!htb]
\centering\small
\caption{Schedules compared at the same post-attention input and prefix-inclusive output boundary. The two Mega MoE rows describe the benchmark wrappers.}
\label{tab:baselines}
\begin{tabular}{@{}L{.16\linewidth}L{.24\linewidth}L{.27\linewidth}L{.24\linewidth}@{}}
\toprule
Path & Front & Routed and shared experts & Output \\
\midrule
SGLang-style & Merged gate/down, then Top16 and FP8 staging & Mega MoE routed call, shared MLP on a side stream & Routed norm and up projection, then join and add shared/prefix \\
vLLM-style & Gate/Top16 can overlap latent down projection, then staging & Mega MoE routed call, then shared MLP on the main stream & Routed norm, BF16 projection with shared addend, then prefix add \\
\Kthree{} & Joint front projection with latent quantization, then router & Dispatch and a four-phase expert compute scheduler on separate SM partitions & Peer-ready combine, norm, up projection and shared/prefix addition \\
\bottomrule
\end{tabular}
\end{table}

\section{The role graph}
\label{sec:graph}

We describe the implementation at commit \texttt{53f430914ea8f3de196d9a53ba8b9750ca6ba8d6} \cite{gck3code}. For a positive physical row count $T$, the base roles are Front (1), Router (2), Dispatch (3), Compute (4), and Finalize (5). The runtime rounds the compute SM request down to the device's co-scheduling alignment and obtains the partition through the CUDA driver. The remaining SMs form the communication resource. The default dispatch \emph{request} is eight SMs through EP8 and sixteen at larger EP counts. The aligned split determines the actual SM counts. The implementation uses balanced workqueue resources where supported \cite{green}. Two partitions permit concurrent progress. Physical concurrency also depends on the device's workqueues, residency, and the instructions issued by both partitions.

\begin{figure}[htbp]
  \centering
  \includegraphics[width=\linewidth]{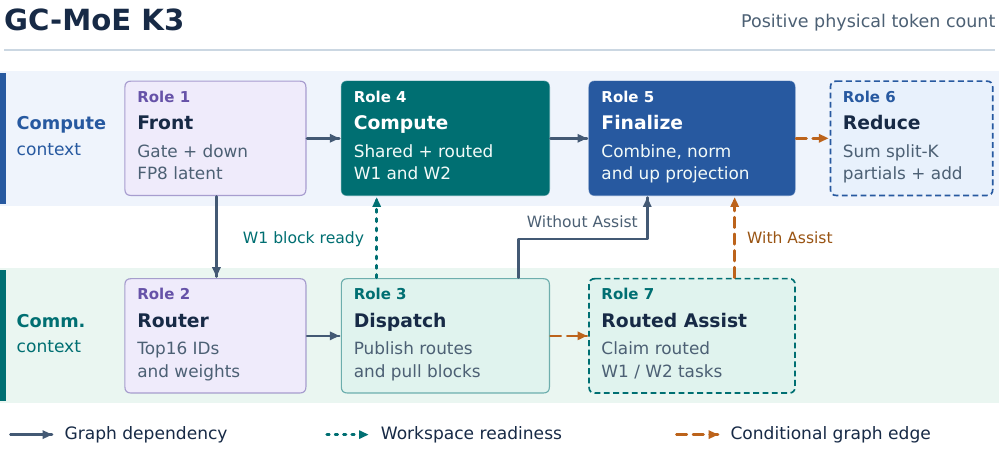}
  \caption{GC-MoE K3 dependencies across two Green Contexts. Front normally completes before Compute starts. The early-shared plan can trigger Compute sooner. The dotted edge carries input-block readiness to routed W1. Shared W1 can proceed independently. Without Assist, Finalize joins Dispatch and Compute. With Assist, the path through Role 7 replaces the direct Dispatch-to-Finalize edge. Split-$K$ projection enables Role 6.}
  \label{fig:roles}
\end{figure}

On the Green Context path, Front, Compute, Finalize, and the optional projection Reduce use the compute context. Router, Dispatch, and the optional Routed Assist use the communication context. Figure~\ref{fig:roles} shows the dependency graph. Dispatch always follows Router. Compute normally follows Front and can enter shared work independently of Router. When the early-shared plan is selected, the Front-to-Compute edge is programmatic and Front triggers the dependent launch early. The routed part of Compute is still gated by published metadata and ring data. Finalize joins local Dispatch and Compute completion. With an admitted Routed Assist node, Finalize joins Assist and Compute instead, since Assist already follows Dispatch. Projection Reduce is a successor of Finalize when the up projection uses two $K$ partitions. A primary-context diagnostic mode uses a Router-to-Compute completion edge and does not offer the Green early-shared or Assist variants.

A zero-physical-row rank runs Dispatch, Compute, and Finalize to serve remote tokens and retire its workspace. It has no local Front, Router or projection work. For a fixed positive physical graph shape and a smaller logical live count, Router marks out-of-prefix routes invalid. The operator defines outputs for live rows. These cases distinguish an empty source rank from an EP group with no work anywhere.

\section{Role implementation and overlap}
\label{sec:implementation}

\subsection{Roles 1 and 2: combined front projection and route publication}

Front reads $X$ and a BF16 matrix of shape $[896+3584,7168]$ containing the router and down-projection weights. Two thread blocks (CTAs) form a cluster. They stage operands with the Tensor Memory Accelerator (TMA), issue BF16 tensor-core operations, and accumulate in tensor memory (TMEM). The epilogue stores gate columns as FP32 logits. For the latent columns it rounds to BF16 and immediately writes E4M3 activations and one UE8M0 exponent per 32 elements to the symmetric workspace. The scale is the group absolute maximum clamped to at least $10^{-10}$, divided by 448 and rounded up to a power of two. This avoids publishing a separate dense latent tensor before staging. The original input remains available to the independent shared branch.

Router assigns one 256-thread CTA per physical row. Threads scan panels of logits, compute the sigmoid scores with correction bias, and select warp candidates. A final warp merges the candidates. It resolves ties in favor of the smaller ID, writes winners in increasing expert-ID order, and normalizes their unbiased sigmoid scores when route renormalization is enabled, as in the benchmark. The optional valid-token scalar gates this selection. Masked rows write the invalid ID and zero weight. Dispatch reads only valid routes.

The planner derives early-shared eligibility from launch geometry. It reserves cluster capacity for Front and checks that the remainder can host at least half of one shared W1 row tile's tasks, capped by the compute cluster count. With more $K$ slices than twice the pipeline depth, Front can trigger Compute after issuing two rounds of TMA loads. Shared W1 can start with $X$ and local shared weights. Compute still waits on the Front dependency before returning. The geometry allows a partial shared wave to overlap Front.

\subsection{Role 3: route metadata and block-wise payload exchange}

Each source rank counts its local routes and writes token-slot indices to disjoint destination spans. The single-CTA path counts and assigns slots in one traversal. The EP group shards a contiguous expert-ID range so an expert ID identifies both an owner rank and its local expert index. For groups up to 32 ranks, including all measured configurations, a source publishes its counts and indices with an epoch-tagged readiness signal. The receiver acquires the sources and publishes completed per-expert totals locally. Small eligible shapes use a single CTA for the source route traversal, avoiding the multi-CTA traversal's grid coordination. For larger groups the implementation also has a per-expert system-atomic publication path.

The destination uses the counts to assign logical blocks of rows to experts. Blocks are padded at expert boundaries and mapped onto a bounded activation ring. Loader warps pull token bytes in stages through shared memory and copy the scale words and source rank/token/slot metadata. After storing the payload and metadata, Dispatch increments the ready count for that routed W1 block with release semantics. The final token in a partial block accounts for the unused rows in that count. A compute worker can start on complete blocks while Dispatch is still populating other blocks. On ring wrap, generation counters prevent overwrites. Dispatch waits for W1 to release the input slot, and W1 waits for W2 to release the intermediate slot.

\subsection{Role 4: shared and routed computation in one task schedule}

An expert task identifies an M block, N tile, layer phase, local expert, and valid row count. The lead CTA's scheduler writes the 32-byte descriptor to the same alternating slot in each CTA's shared memory. The first warpgroup contains the scheduler, TMA load and MMA workers. Two further warpgroups implement the epilogue. Transaction barriers manage shared-memory operand stages and TMEM accumulator handoffs. Routed computation uses operand-swapped SM100 block-scaled \texttt{tcgen05} MMA for FP8 activations and FP4 weights. The shared tasks instead run BF16 MMA without microscale operands. Offline transforms interleave W1 gate/up rows and pack W1 and W2 payloads by N and K slice. FP4 payload slices span 128 K elements, BF16 payload slices span 64. The packed weight scales separately follow the tensor-core scale layout. The mixed-format instruction consumes FP4 weights as operand A and FP8 activations as operand B. Scale tiles move into TMEM before the block-scaled instructions.

The scheduler can issue shared W1 before routed input blocks are ready. Here W1 computes the gate and up projections, and W2 maps the gated activation back to the branch output width. Depending on the shape, it can issue shared W2 immediately after shared W1, after an initial routed W1 task, or after the routed traversal. At small eligible shared shapes it can queue shared W2 while shared W1 is producing its inputs. W2 load stages wait for the matching W1 K64 slice, while other shapes use a coarser readiness count. Routed W1 rounds the gate/up results to BF16, evaluates SiTU, and writes K32 FP8 intermediate data. W2 can begin consuming ready slices of an intermediate block. Each K128 load waits for the two relevant W1 completion bits, with the expected mask adjusted for the ring generation. When the live routed set fits into the ring, the scheduler can batch the W1 traversal. With a wrapping ring it interleaves W1 and W2 work so the latter releases slots. Routed W2 scatters BF16 values with aligned vector stores to the originating token's rank. Each returned slot holds a 3584-element row, and its address is computed from the top-k slot, source row, and channel offset. Shared W2 writes a full-width BF16 result locally.

\paragraph{Role 7: conditional Routed Assist.}
The planner considers Assist when the communication partition is suitable for two-CTA groups and route geometry offers additional work. The runtime checks whether at least one cluster can reside under the actual communication-context launch configuration. An admitted Assist starts after Dispatch and claims the routed W1/W2 tasks from the same queues as Compute. It performs no shared MLP tasks and does not claim work if the live routed set would wrap the ring. This reuses the communication SMs after their primary job while avoiding a change to ring progress conditions. The five-role graph remains valid if the admission check rejects Assist.

\subsection{Roles 5 and 6: peer-ready finalization and up projection}

After the local workers complete, an elected CTA in the first pair to claim publication broadcasts the rank's return epoch to all peers. The receiver checks that epoch only for the owners of a token's valid expert slots. All tokens selecting the same owner check the same publication. Combine warps load the BF16 responses by slot, multiply by the stored route weights, and accumulate in FP32 in a fixed slot order. They round the latent sum to BF16. A nonunit routed scale multiplies that sum and introduces a second BF16 rounding before RMSNorm. The normalized latent row is written in BF16 and released with a row-ready signal. Invalid slots make no contribution.

Two-CTA BF16 projection workers prefill weight stages while combine warps produce rows. The projection issues latent TMA loads once the matching rows are ready and computes the $3584\to7168$ expansion from those normalized rows. A two-slot accumulator pipeline allows the next projection tile to use TMEM as the epilogue finishes stores. The epilogue adds the local shared output. When the selected tile geometry and available partial buffer favor split-$K$, Finalize writes two FP32 projection partials. Role 6 sums them and performs the same shared and optional prefix addition. It adds an output dependency, represented by the final completion edge in Figure~\ref{fig:roles}.

For prefix-inclusive output, GC rounds the FP32 projection and performs the two BF16 additions in the order $\bfround(\bfround(\bfround(U)+S)+P)$. Without a prefix it adds $S$ to the FP32 projection before rounding. These boundaries matter for comparisons with the vLLM-style beta-add and separate prefix add. In the measured Mega MoE configurations, the W1 epilogue applies the route weight to the SiTU output before quantizing it for W2. In GC it multiplies the returned expert contribution in the Finalize reduction. A linear W2 permits either location in real arithmetic. Quantization and rounding make the finite-precision computations distinct. The evaluation compares all paths against corresponding numerical references.

\subsection{Workspace replay and ordering}

Separate mechanisms handle graph completion and data visibility. Local graph edges order role launches and joins. Release/acquire handshakes publish the dispatch metadata, ring readiness, returned peer epochs, and normalized rows. The workspace uses two receive-count generations. Finalize advances a persistent parity and clears the local task and ring state used by the completed call. The subsequent Dispatch clears the retired receive-count bank before publishing the new generation. Completion of the count protocol, including zero-count experts, orders future reuse. Peer-return epochs protect the per-token contribution reads and distinguish stale publications. This protocol removes a trailing cross-rank workspace barrier while preserving communication in Dispatch and peer waits in Finalize.

The runtime caches compiled role binaries, TensorMaps and generated child graphs. During capture it attaches the child graph to the caller's graph. Eager launches instantiate a cached child graph. Captured graph objects retain storage used by their kernels. Calls and replays that share the same workspace allocation must be serialized. These mechanisms apply to the supplied graph and shape configuration. We use whole-operator measurements for the aggregate latency comparison.

\section{Evaluation}
\label{sec:eval}

\subsection{Setup and provenance}

The benchmark report contains measurements on NVIDIA GB200 (reported device name L20A) at EP8, EP16, and EP32 \cite{gck3results}. For each EP size we use symmetric token counts of 1, 2, 4, 8, 16, 32, 64, and 128 per rank. The harness generates weights, hidden states, and a per-token residual addend with fixed seeds and Kimi K3 shapes. Routed weights use FP4 and routed activations use FP8 with scales for groups of 32 elements. The shared MLP and full-width projection use BF16 weights. The measured interval covers from ready post-attention hidden states through the final output including the prefix. Attention, request handling, weight transformations, workspace allocation and first graph capture are outside the timed interval.

The benchmark captures one operator invocation. On each rank it divides the CUDA event time of each window by the number of replays in that window, then takes the median of those per-replay estimates. The reported latency is the maximum of the rank medians. The benchmark defaults are ten warmup windows, 30 measured windows, and 20 replays per window. Barriers, cache eviction and synchronization preceding each timed window are outside the event interval. The source alternates the two timing orders across token counts.

\paragraph{Paired and historical data.}
The 24 GC and SGLang values are the same-round measurements from the benchmark report under PyTorch 2.13. The reported vLLM values are historical measurements carried forward from an earlier PyTorch 2.14 run. Both ratios use the current GC latency as denominator. The comparison with vLLM therefore spans two software environments.

\subsection{Measured latency}

\begin{figure}[!htb]
  \centering
  \includegraphics[width=\linewidth]{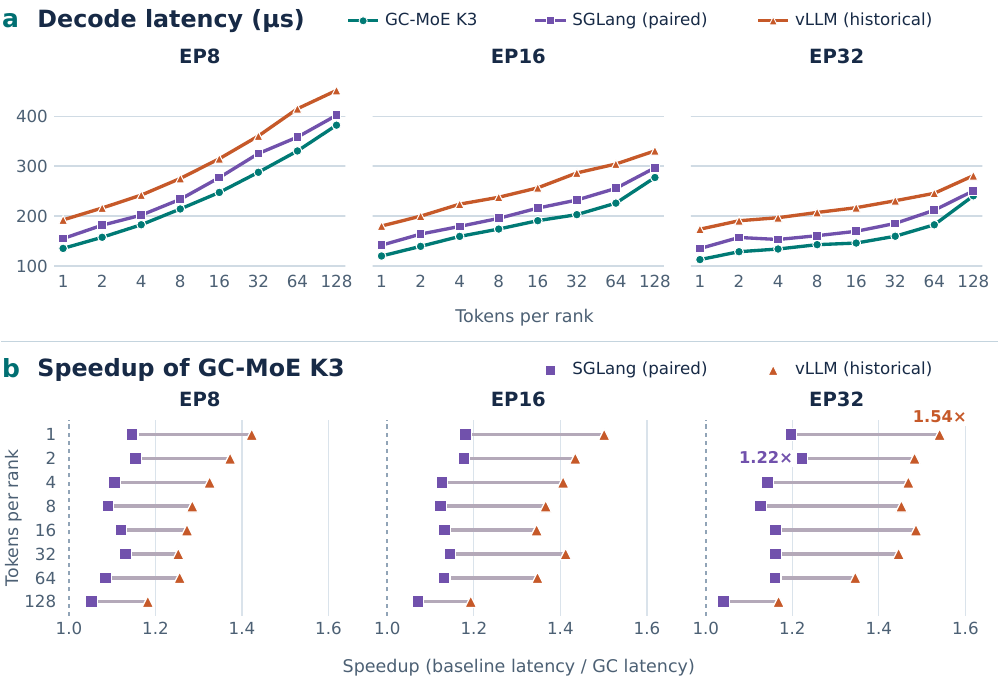}
  \caption{GB200 decode at 24 EP/token configurations. \textbf{a,} Lines join the measured maximum of rank-median latencies at successive token counts. \textbf{b,} Markers show the baseline/GC latency ratios for the same EP size and token count. The dashed line marks parity ($1\times$), and labels mark each baseline's peak ratio. GC and SGLang share a measurement run. vLLM uses historical measurements.}
  \label{fig:latency}
\end{figure}

\begin{table}[htbp]
  \centering\small
  \caption{Speedup ratios from measured latencies. Each mean is an unweighted mean of per-point baseline/GC ratios. GC is faster at every listed point against both comparison series.}
  \label{tab:speedup}
  \begin{tabular}{@{}lrrrrr@{}}
  \toprule
   & & \multicolumn{2}{c}{SGLang / GC (paired)} & \multicolumn{2}{c}{vLLM / GC (historical)} \\
  \cmidrule(lr){3-4}\cmidrule(l){5-6}
  Group & Points & Mean & Peak & Mean & Peak \\
  \midrule
  EP8 & 8 & 1.1102 & 1.1537 & 1.2962 & 1.4226 \\
EP16 & 8 & 1.1354 & 1.1804 & 1.3756 & 1.5011 \\
EP32 & 8 & 1.1514 & 1.2222 & 1.4236 & 1.5404 \\
All & 24 & 1.1324 & 1.2222 & 1.3651 & 1.5404 \\
\bottomrule

  \end{tabular}
\end{table}

The paired SGLang speedups average $1.1102\times$, $1.1354\times$, and $1.1514\times$ at EP8, EP16, and EP32, respectively (Table~\ref{tab:speedup}). The peak paired speedup is $1.2222\times$ at EP32 with two tokens per rank (128.456 vs.\ 156.994~\US{}). At EP32 with 128 tokens per rank it narrows to $1.0413\times$ (240.379 vs.\ 250.310~\US{}). The historical vLLM ratio peaks at $1.5404\times$ at EP32 with one token per rank. Across all points the respective means are $1.1324\times$ and $1.3651\times$. Changing EP at fixed tokens per rank changes both the local expert count and the global token count.

\subsection{Numerical accuracy}

Each comparison suite contains default, boundary, and asymmetric rank inputs, including zero-row ranks. The EP8, EP16, and EP32 suites cover 37, 45, and 60 cases. The checker exercises graph replay and eager calls with and without a prefix. The references decode quantized operands and model each path's intermediate rounding and output addition order. In the vLLM-style RMSNorm, the normalized value is rounded to BF16 before applying the norm weight. GC applies the norm weight before rounding the normalized result. The acceptance thresholds are at least 0.99 cosine similarity and at most 0.02 relative $\ell_2$ error to the corresponding reference, with the same cosine threshold for the direct comparison between operators. The checker also verifies that masked rows have invalid expert IDs and zero route weights. The paired GC/SGLang experiment passes all 142 cases and 2{,}936 rank checks. The historical GC/vLLM suite also passes that many. Table~\ref{tab:accuracy} summarizes the worst reported reference comparisons across the respective suites.

\begin{table}[htbp]
  \centering\small
  \caption{Numerical results against each path's reference. GC and SGLang come from the paired experiment. vLLM is historical.}
  \label{tab:accuracy}
  \begin{tabular}{@{}lrrr@{}}
  \toprule
  Path & Minimum cosine & Maximum relative $\ell_2$ & Maximum absolute error \\
  \midrule
  GC-MoE K3 & 0.999995410 & 0.003025480 & 0.50 \\
  SGLang-style & 0.999996305 & 0.002748258 & 0.50 \\
  vLLM-style & 0.999996006 & 0.002818683 & 0.50 \\
  \bottomrule
  \end{tabular}
\end{table}

\paragraph{Limits.}
The synthetic inputs and weights can yield different expert skew and cache behavior from deployed checkpoints. The latency results cover symmetric inputs at the operator boundary. Asymmetric inputs are used for numerical validation. Individual role effects and serving performance would require matched ablations and system-level measurements.

\section{Related work}

GC-MoE introduced Green-Context separation of token dispatch and grouped GEMM for ordinary expert-parallel MoE \cite{gcbase}. Its SM120 extension adapts the pipeline to warp-level matrix instructions and PCIe \cite{gcsm120}. This paper targets the SM100 family and K3's wider operator boundary, in which front projection and routed normalization and projection have material dependencies outside the expert MLPs. The SGLang and vLLM paths use the Mega MoE approach to combine the routed exchange and expert MLPs within a device call \cite{sglang,sgldg,vllm,vllmdg}. Both framework paths schedule the shared MLP and the output projection outside that call. GC schedules them in the same role graph as the routed computation.

\begin{samepage}
\section{Conclusion}

\Kthree{} schedules the Kimi K3 sublayer from its post-attention input through routing, expert computation, and the normalized projected sum. Dispatch publishes ready input blocks to Compute. After the expert results have been written, each owner rank publishes one return epoch for Finalize to check. The scheduler can overlap shared W1 with front and communication work and reuse the dispatch partition after token exchange. Finalize combines routed contributions and prepares rows for the BF16 up projection using token-specific peer waits. On the GB200 sweep, \Kthree{} improves on the paired SGLang-style path at all 24 measured points. Its mean paired speedup is $1.1324\times$ at the evaluated prefix-inclusive boundary.

\end{samepage}

\bibliographystyle{unsrt}
\bibliography{refs}

\begin{thebibliography}{1}

\bibitem{gcbase}
Yong Li, Kan Liu, and Tao Lan.
\newblock {GC-MoE}: Overlapping expert-parallel token dispatch with {FP8/FP4} grouped {GEMM} via {CUDA Green Contexts}.
\newblock arXiv:2611.04222, 2026.
\newblock \url{https://arxiv.org/abs/2611.04222}.

\bibitem{gcsm120}
Yong Li, Haoyu Wang, Kan Liu, and Tao Lan.
\newblock {GC-MoE} on {SM120}: Overlapping expert-parallel token dispatch with warp-level {FP8xFP4} block-scaled {MMA} on consumer {Blackwell}.
\newblock arXiv:2611.16219, 2026.
\newblock \url{https://arxiv.org/abs/2611.16219}.

\bibitem{sglang}
{SGLang contributors}.
\newblock {Kimi K3} model implementation in {SGLang}, 2026.
\newblock Revision da395ae091. \href{https://github.com/sgl-project/sglang/blob/da395ae0913ab92e57d9c1faeb5ab6b33a914b23/python/sglang/srt/models/kimi_k3.py}{Model source}.

\bibitem{sgldg}
{SGLang contributors}.
\newblock {SGLang DeepGEMM: Mega MoE implementation}, 2026.
\newblock Version 0.2.8. \href{https://github.com/sgl-project/DeepGEMM/tree/e3f6b84093dab1213f50cd1865d847365697ef23}{Pinned source}.

\bibitem{vllm}
{vLLM contributors}.
\newblock {Kimi K3} {NVIDIA} model implementation in {vLLM}, 2026.
\newblock Revision 89d9b4d187. \href{https://github.com/vllm-project/vllm/blob/89d9b4d187efbe03a9a4fee72d668587a4629d18/vllm/models/kimi_k3/nvidia/model.py}{Model source}.

\bibitem{vllmdg}
{vLLM contributors}.
\newblock {DeepGEMM fork used by vLLM: SM100 SiTU Mega MoE}, 2026.
\newblock Revision 44cba7d. \href{https://github.com/vllm-project/DeepGEMM/tree/44cba7d28066182852005c59052f035ebc18e211}{Pinned source}.

\bibitem{gck3code}
{GC-MoE contributors}.
\newblock {GC-MoE K3} source and evaluation harness, 2026.
\newblock Implementation revision 53f4309. \href{https://github.com/Bruce-Lee-LY/gc-moe}{Project repository}.

\bibitem{green}
{NVIDIA Corporation}.
\newblock {CUDA Driver API}: {Green Contexts}, 2026.
\newblock \href{https://docs.nvidia.com/cuda/cuda-driver-api/cuda_driver_api/group__CUDA__GREEN__CONTEXTS.html}{Driver API reference}.

\bibitem{gck3results}
{GC-MoE contributors}.
\newblock {GB200 Kimi K3} decode latency and numerical accuracy at {EP8, EP16, and EP32}, 2026.
\newblock Test report at revision 53f4309. The source bundle includes the original report under \texttt{anc/data/}. See \hyperref[apx:latency]{Appendix~\ref*{apx:latency}} for the measured latencies.

\end{thebibliography}

\clearpage
\appendix
\section{Measured latency values}
\label{apx:latency}

Table~\ref{tab:latency} lists the measured delays used in Figure~\ref{fig:latency}. The source bundle includes the original benchmark report and the script that regenerates Figure~\ref{fig:latency} and the latency tables from its full-precision entries. Table entries are shown to three decimal places in microseconds.

\begin{table}[!htb]
  \centering\small
  \caption{Maximum of rank-median latency (\US{}). GC and SGLang are same-round values under PyTorch 2.13. vLLM values are historical under PyTorch 2.14.}
  \label{tab:latency}
  \begin{tabular}{@{}rrrrrr@{}}
  \toprule
  EP & Tokens/rank & Total tokens & GC-MoE K3 & SGLang (paired) & vLLM (historical) \\
  \midrule
  8 & 1 & 8 & 135.065 & 154.775 & 192.141 \\
8 & 2 & 16 & 157.357 & 181.541 & 215.983 \\
8 & 4 & 32 & 182.449 & 201.645 & 241.794 \\
8 & 8 & 64 & 214.038 & 233.386 & 275.104 \\
8 & 16 & 128 & 247.180 & 276.877 & 314.660 \\
8 & 32 & 256 & 287.506 & 325.036 & 360.202 \\
8 & 64 & 512 & 330.219 & 357.984 & 414.777 \\
8 & 128 & 1024 & 381.850 & 401.629 & 451.474 \\
\midrule
16 & 1 & 16 & 119.879 & 141.504 & 179.954 \\
16 & 2 & 32 & 139.171 & 163.740 & 199.621 \\
16 & 4 & 64 & 159.087 & 179.103 & 223.705 \\
16 & 8 & 128 & 173.848 & 195.220 & 237.470 \\
16 & 16 & 256 & 190.626 & 215.784 & 256.377 \\
16 & 32 & 512 & 202.691 & 231.941 & 286.224 \\
16 & 64 & 1024 & 225.721 & 255.298 & 304.059 \\
16 & 128 & 2048 & 277.039 & 296.546 & 330.424 \\
\midrule
32 & 1 & 32 & 112.664 & 134.846 & 173.552 \\
32 & 2 & 64 & 128.456 & 156.994 & 190.450 \\
32 & 4 & 128 & 133.874 & 152.922 & 196.571 \\
32 & 8 & 256 & 142.553 & 160.566 & 207.027 \\
32 & 16 & 512 & 145.797 & 169.329 & 216.627 \\
32 & 32 & 1024 & 159.448 & 185.091 & 230.522 \\
32 & 64 & 2048 & 182.385 & 211.616 & 245.426 \\
32 & 128 & 4096 & 240.379 & 250.310 & 280.806 \\
\bottomrule

  \end{tabular}
\end{table}

\section{Reproducing the workload}

The commands below require the implementation revision identified in Section~\ref{sec:graph}. The script \texttt{scripts/run\_k3.sh} selects the Kimi K3 shape and the default eight token counts. Run \verb|bash scripts/run_k3.sh 8 both --baseline sglang| in the SGLang dependency environment. Substitute 16 or 32 and set the script's node/rank configuration for the desired EP group. The SGLang wrapper requires \texttt{sgl-deep-gemm} 0.2.8 and uses FlashInfer for GEMM and RMSNorm. For vLLM use \verb|--baseline vllm| with the DeepGEMM revision cited in Section~\ref{sec:baselines}. Use separate dependency environments because both builds export \texttt{deep\_gemm} and must match the active PyTorch ABI. Record the gate implementation, package versions, CUDA driver/runtime, node topology, physical SM allocation, and graph shape for any fresh measurements. New matched-environment measurements are needed to remove the historical comparison caveat.

\end{document}